\documentclass[twocolumn,prd,showpacs,10pt]{revtex4}
\usepackage{graphicx}
\usepackage{dcolumn}
\usepackage{bm}
\usepackage{amsmath}
\usepackage{hyperref}

\newcommand{\gsim}{\gtrsim}
\newcommand{\lsim}{\lesssim}
\newcommand{\square}{\kern1pt\vbox{\hrule height 1.2pt\hbox{\vrule
width 1.2pt\hskip 3pt
\vbox{\vskip 6pt}\hskip 3pt\vrule width 0.6pt}\hrule
height 0.6pt}\kern1pt}
\newcommand{\beq}{\begin{equation}}
\newcommand{\beqn}{\begin{eqnarray}}
\newcommand{\eeq}{\end{equation}}
\newcommand{\eeqn}{\end{eqnarray}}

\begin{document}

\title{Primordial Non-Gaussianity and Gravitational Waves: 
Observational Tests of Brane Inflation in String Theory}
\author{James E. Lidsey and David Seery}
\affiliation{Astronomy Unit, School of Mathematical Sciences, 
Queen Mary, University of London, Mile End Road,
London, E1 4NS, UK}

\begin{abstract}
We study brane inflation scenarios in a warped throat geometry and show 
that there exists a consistency condition between the non-Gaussianity
of the curvature perturbation and the amplitude and scale-dependence 
of the primordial gravitational waves. This condition is 
independent of the warping of the throat and the form of the 
inflaton potential. We find that such a relation could be tested by a 
future CMB polarization experiment if the Planck satellite is able to 
detect both a gravitational wave background and a non-Gaussian statistic. 
In models where the observable stage of inflation occurs when the 
brane is in the tip region of the throat, we derive a further consistency 
condition involving the scalar spectral index, the tensor-scalar ratio and 
the curvature perturbation bispectrum.     
We show that when such a relation is combined with the 
WMAP3 results, it leads to a model-independent 
bound on the gravitational wave amplitude given by  
$10^{-3} < r < 10^{-2}$. This corresponds to the 
range of sensitivity of the next generation of
CMB polarization experiments. 
\end{abstract}

\vskip 1pc 
\maketitle

\section{Introduction}
  
The inflationary scenario, whereby the universe underwent a phase 
of accelerated expansion in its most distant past, represents the 
cornerstone of modern, early universe cosmology 
\cite{simplest,perturbations}. It has proved remarkably 
successful when confronted with cosmological observations, 
in particular the three-year data from the 
Wilkinson Microwave Anisotropy Probe (WMAP3) \cite{spergel}.  
String theory is presently the favoured candidate for a unified theory of 
the fundamental interactions including gravity and it is therefore important 
to embed inflation within string theory. 

One approach to string-theoretic 
inflation is based on ${\rm D}$-branes 
\cite{earlybrane,kklt,kklmmt,silverstein,chensolo,shandtye,tipinflation,otherbranes}.
Of particular interest are scenarios where a type IIB orientifold is 
compactified on a Calabi-Yau (CY) three-fold,  
where the moduli fields are stabilized 
due to the presence of non-trivial flux \cite{gkp,kklt}. 
(See \cite{fluxcompact} for a review). These fluxes generate  
local regions within the CY space with a warped geometry or `throat'. 
In many settings, an anti-${\rm D3}$-brane 
sits naturally at the infra-red (IR) tip of the throat and 
attracts a ${\rm D3}$-brane towards it. 
The brane separation then plays the role of the inflaton 
and, since this is an open string mode, 
its dynamics is determined by a Dirac-Born-Infeld (DBI) action.  
In general, there are two limiting 
regimes for brane inflation which are characterized by the 
rolling of the inflaton. In the KKLMMT scenario, for example, 
the inflaton undergoes conventional slow-roll down a flat potential
\cite{kklmmt}. However, the non-linear nature of the 
DBI action implies that inflation may also proceed when the field 
is rolling relatively fast, as is the case in DBI inflation 
\cite{silverstein,chensolo,shandtye}. 

In this paper, we investigate how the brane inflationary scenario could 
be tested if forthcoming observations of the Cosmic Microwave Background (CMB)
uncover non-Gaussian statistics in the scalar (density) perturbation spectrum
and also a primordial background of tensor (gravitational wave) perturbations. 
An observed departure from purely Gaussian statistics will provide a 
powerful discriminant between different inflationary models and the 
Planck satellite, which is scheduled for launch in 2007, will 
improve on the WMAP3 sensitivity by an order of magnitude \cite{planck}. 
On the other hand, the detection of gravitational waves 
will open up a direct window onto the energy scale of inflation. 
The most promising means of detecting tensor perturbations is 
through the $B$-mode polarization of the CMB and a number of 
experiments, such as Clover (the `Cl-Observer') \cite{clover}, are 
presently under construction. (For a review, see, e.g., \cite{cmbreview}). 

The paper is organized as follows. We begin by summarizing the 
brane inflation scenario in Section 2. In Section 3, we derive 
a consistency condition between the non-Gaussianity 
of the curvature perturbation and the amplitude and scale-dependence of the 
gravitational waves. This relation is independent of 
the warped geometry and the inflaton potential. 
We then discuss the prospects for testing such a condition with 
future CMB polarization experiments. In Section 4, we consider the class of 
models where the observable phase of inflation 
occurred when the ${\rm D}$-brane was near the IR tip of the throat. 
We derive a further consistency condition in this regime 
that relates the scalar spectral index, the tensor-scalar ratio and 
the bispectrum of the curvature perturbation. We find that such a 
constraint leads to a bound on the gravitational wave amplitude of 
$10^{-3} < r < 10^{-2}$, which is in the 
range of sensitivity of the next generation of 
CMB polarization experiments. We conclude with a discussion in Section 6. 

\section{Brane Inflation in String Theory}

In general, the low energy world-volume dynamics
of a probe ${\rm D3}$-brane in a warped background is given by 
\begin{eqnarray}
\label{action}
S=\int  d^4x \sqrt{|g|} \left[ \frac{m_{\rm pl}^2}{16\pi} R 
+ P (\phi , X) \right] 
\\
\label{defP}
P( \phi ,X) = f_1(\phi ) \sqrt{1+f_2 (\phi ) X}
-f_3 (\phi )   ,
\end{eqnarray}
where $R$ is the Ricci curvature scalar,
$X \equiv - \frac{1}{2} g^{\mu\nu} \nabla_{\mu} \phi \nabla_{\nu} \phi$
is the kinetic energy of the inflaton field, $\phi$, 
and the functions $f_i (\phi )$ are 
determined by the warped geometry and the inflaton self-interaction 
potential. 

A concrete example of a warped background is  
the Klebanov-Strassler (KS) solution of the type IIB theory, 
where the throat is a warped deformed conifold \cite{ks}. 
The ten-dimensional metric has the form 
$ds_{10}^2= \tilde{f}^{-1/2} (\tau) \eta_{\mu\nu} dx^{\mu} dx^{\nu}
+\tilde{f}^{1/2} (\tau ) ds_6^2$, where $\tilde{f}(\tau)$ denotes 
the warp factor and $\tau$ is the coordinate 
along the throat (corresponding to one of the coordinates 
on $ds_6^2$). In the ultra-violet (UV) regime ($\tau \rightarrow 
\infty$) the throat corresponds to a cone 
over the Einstein manifold $T^{1,1}= {\rm SU(2)} \otimes {\rm SU(2)}/
{\rm U(1)}$, which has a topology $S^2 \times S^3$, where the $S^2$ is 
fibred over the $S^3$. On the other hand, at the tip of the throat ($\tau 
\rightarrow 0$) the wrapping of the fluxes along the cycles of the conifold
smooths out the conical singularity with an $S^3$ `cap' \cite{ks,kt}. 

The effective DBI action for a ${\rm D}$-brane in this background 
is given by \cite{tipinflation}
\begin{equation}
P = -T_3 \left[ \frac{1}{\tilde{f} (\tau)} 
\left( \sqrt{1+ h(\tau ) (\nabla \tau )^2} -1 \right) +V(\tau ) \right]  ,
\end{equation}
where the warp factor is 
\begin{eqnarray}
\label{warp}
\tilde{f} (\tau ) = A I(\tau ) , \qquad 
h(\tau ) = B \frac{I(\tau )}{K^2(\tau )}
\\
\label{defI}
I(\tau ) = \int_{\tau}^{\infty} dx \, \frac{x \,{\rm coth} (x) -1}{\sinh^2 (x)}
\left[ \sinh (2x) -2x \right]^{1/3}
\\
\label{defK}
K (\tau ) = \frac{[\sinh (2\tau ) - 2\tau ]^{1/3}}{2^{1/3} \sinh (\tau )}  .
\end{eqnarray}
The brane tension is denoted by $T_3$, 
$V(\tau )$ is a non-perturbative potential
and $\{ A,B \}$ are model-dependent constants.  

Since the aim of this paper is to identify model-independent, observational 
tests of brane inflation that are insensitive to 
the specific form of the inflaton potential and the nature of the warped 
geometry, we will assume that 
the form of the kinetic function $P (\phi , X)$ is given by Eq. (\ref{defP}) 
for arbitrary $f_i (\phi)$, subject only to the condition that 
a successful phase of inflation can be realised.
We will exploit the fact that the non-trivial kinetic structure 
in $P (\phi , X)$ causes the sound speed of fluctuations 
in the inflaton to differ from unity (the value 
it takes in canonical, single-field inflation, where 
$P=X - V(\phi)$). This will lead to   
observational signatures involving both non-Gaussianity 
and gravitational waves. 

The Friedmann equations derived from action (\ref{action}) take the form
\begin{equation}
\label{friedmann}
3m_{\rm pl}^2H^2 = 8\pi E , \qquad \dot{E} = -3H (E+P)  ,
\end{equation}
where $E=2XP_{,X} -P$, $H$ represents the Hubble parameter, a dot
denotes $d/dt$ and a comma denotes a partial derivative. 
The inflationary dynamics for this class of models 
can be quantified in terms of three `slow-roll' parameters:
\begin{equation}
\label{defslowroll}
\epsilon \equiv - \frac{\dot{H}}{H^2}, \qquad 
\eta \equiv \frac{\dot{\epsilon}}{\epsilon H} , \quad 
s \equiv \frac{\dot{c}_s}{c_sH}   ,
\end{equation}
where 
\begin{equation}
\label{defcs}
c_s^2 \equiv  \frac{P_{,X}}{P_{,X} +2XP_{,XX}}
\end{equation}
defines the sound speed, $c_s$. 
The rate of change of the sound speed is determined by $s$ and
we will assume throughout that $\{ \epsilon, |\eta | , |s | \}  \ll 1$
during the observable phase of inflation.   

The amplitudes of the scalar and tensor perturbation spectra are given by 
\begin{equation}
\label{amplitudes}
P_S^2= \frac{1}{\pi m_{\rm pl}^2} \frac{H^2}{c_s \epsilon}
, \qquad
P_T^2 = \frac{16}{\pi} \frac{H^2}{m_{\rm pl}^2} ,
\end{equation} 
respectively, and the corresponding spectral indices are \cite{gm}
\begin{equation}
\label{defindices}
1-n_s = 2 \epsilon +\eta +s , \qquad  n_t = -2\epsilon  .
\end{equation}
The tensor-scalar ratio, $r \equiv P_T^2/P_S^2$, is directly related to 
the tensorial spectral index such that \cite{gm}
\begin{equation}
\label{consistency}
r= -8c_sn_t  .
\end{equation}

Finally, deviations from purely Gaussian statistics 
arise when the primordial three-point function for the curvature 
perturbation is non-trivial.  
It is conventional to write the non-Gaussian curvature 
perturbation ${\cal{R}}$ as a sum of a Gaussian part and
the square of a Gaussian such that ${\cal{R}} ={\cal{R}}_G+f_{NL}
({\cal{R}}^2_G -\langle {\cal{R}}^2_G \rangle)$, where 
the quadratic part is a convolution and $f_{NL}$ defines 
the `non-linearity' parameter \cite{maldacena}. 
In general, this parameter is a function of the three momenta which 
form a triangle in Fourier space. Recently, it was 
shown that in the equilateral triangle limit 
(where the three momenta have equal magnitude), 
the leading-order contribution to the non-linearity parameter 
is given by \cite{chen}  
\begin{equation}
\label{deffNL}
f_{NL} = \frac{35}{108} \left( \frac{1}{c_s^2}-1 \right) - 
\frac{5}{81} \left( \frac{1}{c_s^2}-1 - 2\Lambda  \right)  ,
\end{equation}
where 
\begin{equation}
\label{defLambda}
\Lambda \equiv 
\frac{X^2P_{,XX}+\frac{2}{3} X^3P_{,XXX}}{X P_{,X}+2X^2P_{,XX}}  .
\end{equation}
The bound on the non-linearity parameter imposed by 
WMAP3 is
$|f_{NL}| \lsim 300$
\cite{spergel,crim}. 
  
\section{An Observational Test of Brane Inflation}

It was observed in \cite{chen,tipinflation} that 
when $f_1=1/f_2$, the second term on the right-hand side of Eq. (\ref{deffNL}) 
vanishes, i.e.,
\begin{equation}
\label{vanish}
\Lambda  = \frac{1}{2} \left( \frac{1}{c_s^2} -1 \right)  .
\end{equation}
It can be further shown that for arbitrary $f_i(\phi )$, 
Eq. (\ref{defP}) is the most general form of the kinetic function 
$P (\phi , X)$ which satisfies Eq. (\ref{vanish}). Specifically, 
we may employ the 
definitions (\ref{defcs}) and (\ref{defLambda}) to express Eq. (\ref{vanish})  
in the form of a third-order, non-linear partial differential equation: 
\begin{equation}
\label{pde}
P_{,X}P_{,XXX} = 3P^2_{,XX}  .
\end{equation}
However, Eq. (\ref{pde}) can be reduced to the first-order equation 
\begin{equation}
\label{firstorder}
Z_{,X}=2Z^2
\end{equation}
by defining a new variable $Z \equiv P_{,XX}/P_{,X}$. 
This admits the general solution 
$Z (\phi ,X ) = [f(\phi ) -2X]^{-1}$, where $f(\phi )$ is an arbitrary
function. Integrating this solution twice then yields the general solution 
for $P (\phi , X)$, which is precisely the form given in Eq. (\ref{defP}), 
where the $f_i (\phi)$ arise as arbitrary integration functions. 

It follows, therefore, that for the general class of string-theoretic 
models (\ref{defP}), the leading-order contribution to the non-linearity 
parameter is determined entirely by the sound speed: 
\begin{equation}
\label{fnlcs}
f_{NL} = \frac{1}{3} \left( \frac{1}{c_s^2} -1 \right)  .
\end{equation}
This implies that we may substitute Eq. (\ref{fnlcs}) 
into Eq. (\ref{consistency}) to deduce a `consistency' equation 
which is composed entirely of observable parameters: 
\begin{eqnarray}
\label{departure}
r+8n_t = - r_l  r  
\\ 
\label{defrl}
r_l \equiv \sqrt{1+3f_{NL}} -1  .
\end{eqnarray}

\begin{figure}[!t]

\includegraphics[width=9cm]{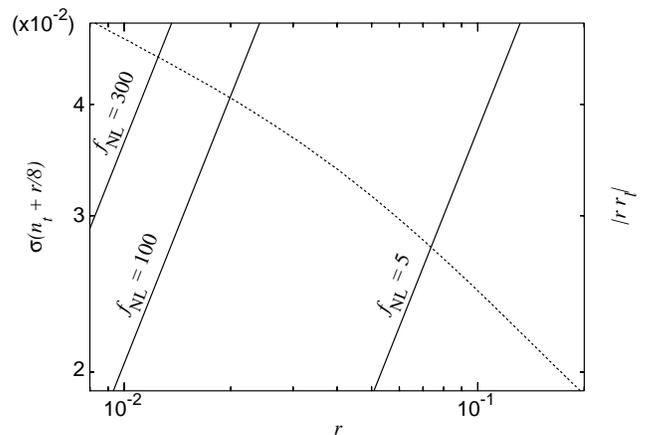}

\caption[] {\label{fig1brane}
The dashed line illustrates the results of Song and Knox 
\cite{SK} anticipating the 
experimental error, $\sigma$, in $r+8n_t$ as a function of $r$ for a
future all-sky CMB polarization experiment with angular resolution
$\varphi = 3'$ and noise $\Delta = 3 \mu{\rm K} \cdot {\rm arcmin}$
\cite{SK}. The modulus of the 
right-hand side of the consistency equation (\ref{departure})
is represented by a straight line for a given value  
of the non-linearity parameter, $f_{NL}$.  
The point of intersection of the lines yields a lower bound on the 
tensor-scalar ratio for a detection of the consistency 
equation (\ref{departure}) to be possible.
}
\end{figure}

\begin{figure}

\includegraphics[width=9cm]{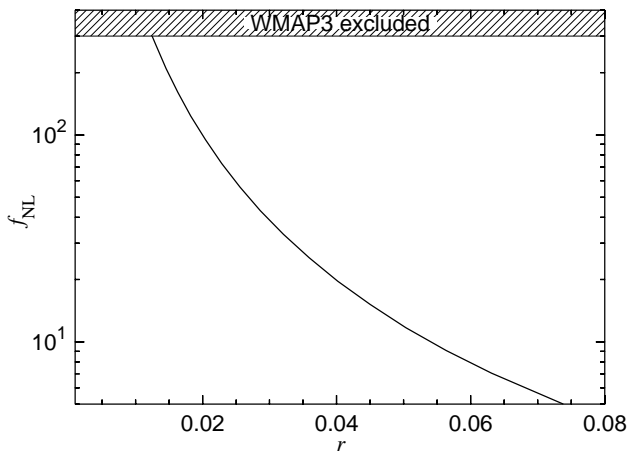}

\caption[] {\label{fig2brane}
Illustrating the minimal value of the non-linearity parameter $f_{NL}$ 
deduced from Fig. \ref{fig1brane} that will lead in principle 
to a detection of the consistency equation (\ref{departure}). 
For a given fiducial value of the tensor-to-scalar ratio, $r$, 
a detection should be possible in the region of parameter 
space above the solid line. The shaded region is excluded by the WMAP3 data. 
}
\end{figure}

Eq. (\ref{departure}) is model-independent in the sense that it holds 
for an arbitrary inflaton potential and a general warping of the 
higher-dimensional spacetime. This is important since the precise form 
of the potential depends on non-perturbative features of the superpotential, 
whereas the warp factor is determined by the details of the 
flux compactification. We may therefore regard 
Eq. (\ref{departure}) as a robust prediction  
for string-theoretic brane inflation 
and it is natural to investigate whether it 
could be employed as a future test of 
the scenario. Since it formally 
reduces to the standard consistency equation of 
slow-roll inflation in the limit $r_l \rightarrow 0$, a detection 
of Eq. (\ref{departure}) would amount to detecting a 
violation of the standard consistency equation. 
A necessary condition for such a detection is that 
the magnitude of the right-hand side of Eq. (\ref{departure})
should exceed the experimental error in the quantity on
the left-hand side \cite{SK,LS}. Recently, 
Song and Knox \cite{SK} have considered 
the level of accuracy that should be attainable with 
future all-sky CMB polarization experiments with angular resolutions 
in the range $1.0' \le \varphi \le 30.0'$ and noise levels between    
$1\mu {\rm K} \cdot {\rm arcmin} \le \Delta \le 
15 \mu{\rm K} \cdot {\rm arcmin}$. They determined the 
anticipated error in $r+8n_t$ as a function 
of the tensor-scalar ratio and this is shown in Fig. \ref{fig1brane}. 
The straight lines in this figure correspond to 
the right-hand side of Eq. (\ref{departure}). For a given 
value of the non-linearity parameter $f_{NL}$, 
the point of intersection between the solid and dashed lines yields 
the minimal value of $r$ above which a detection of Eq. (\ref{departure}) 
will be possible in principle. This bound is illustrated in 
Fig. \ref{fig2brane}. 

It is interesting to consider what could be deduced from the Planck satellite
about our prospects for detecting Eq. (\ref{departure}) 
with a future CMB polarization experiment. Planck will 
have sensitivities down to $r \approx 0.05$ and 
$f_{NL} \approx 5$ \cite{planck,komatsu}. 
For $r = 0.05$, one would require $f_{NL}>12$ for a detection of 
Eq. (\ref{departure}). This is comfortably below the upper limit 
of $f_{NL} \lsim 300$
imposed by the WMAP3 data \cite{spergel}. 
On the other hand, for $f_{NL} \approx 5$, a detection will only 
be possible if $r> 0.07$. Moreover, 
imposing the WMAP3 limit $f_{NL} \simeq 300$ implies a 
future detection would require $r>0.01$, but a value 
$r \approx 0.01$ should be detectable by Clover at 
approximately the $3$-sigma level \cite{clover}. Consequently, 
future all-sky observations will fail to detect Eq. (\ref{departure}) if 
Clover does not detect a primordial tensor spectrum. On the other 
hand, if Planck measures both $r$ and $f_{NL}$, the prospects for 
the detection of Eq. (\ref{departure}) are good.  

\section{Brane Inflation Near the Tip}

Thus far, our discussion has been general, in the sense that 
the functions $f_i (\phi )$ in action (\ref{defP})
have remained unspecified. In this section, we focus on the case where 
the ${\rm D}$-brane is in the vicinity of the IR tip of the warped 
throat. To date, the majority of studies into DBI inflation 
have considered the dynamics of the brane when it is 
far from the tip and in a region where 
the throat is asymptotically ${\rm AdS}_5$
\cite{silverstein,chensolo,shandtye}. However, 
in the UV version of DBI inflation, the brane moves into the 
throat and reheating occurs when it reaches the tip. 
Since the era of inflation that is  accessible to observations 
occurred during the last 30-60 e-folds, the brane may well 
have been near the tip of the throat during that epoch.

DBI inflation in the tip region of the KS background   
was considered recently in Ref. \cite{tipinflation}.
Near the tip, $\tau \rightarrow 0$, and 
it can be shown that 
$I(\tau \rightarrow 0 ) \rightarrow 0.7$ and 
$K(\tau \rightarrow 0) \rightarrow (2/3)^{1/3}$. 
Consequently, the warp factors in Eq. (\ref{warp}) both tend to 
finite constants in this limit \cite{ks}.  

Motivated by the above asymptotic behaviour of the KS throat, 
we consider the class of models (\ref{defP}),  
where $(f_1, f_2) \rightarrow {\rm constants}$. On the other hand, 
we will keep the inflaton potential $f_3 (\phi)$ arbitrary 
(modulo the usual caveats for successful inflation). 
Since $P_{,X\phi}= 0$ in this case, 
it follows after differentiating 
Eq. (\ref{defcs}) with respect to time and using the Friedmann equation 
(\ref{friedmann}) that \cite{SL1}
\begin{equation}
\label{combine}
\Lambda = \frac{2(1+c_s^2)s + (1-c_s^2)(2\epsilon -\eta )}{6c_s^2(2\epsilon -
\eta)}  .
\end{equation}
Combining Eqs. (\ref{vanish}) and (\ref{combine}) 
therefore leads to a constraint equation that relates the sound speed 
directly to the three slow-roll parameters $\{ \epsilon, \eta, s \}$: 
\begin{equation}
\label{equate}
\frac{1+c_s^2}{1-c_s^2} = \frac{2\epsilon -\eta}{s}  .
\end{equation}

Eq. (\ref{equate}) may be converted into a 
consistency relation if four observable parameters involving 
$\{ \epsilon , \eta , c_s , s \}$ can be identified.  
Eqs. (\ref{defindices}) and (\ref{fnlcs}) provide three of these   
in the form of the two 
spectral indices and the non-linearity parameter, $f_{NL}$. 
A possible candidate for the fourth parameter 
is the running of the tensor spectral index, which is defined by 
$\alpha_t \equiv dn_t/d \ln k \approx n_t \eta$, where $k$ 
is the comoving wavenumber. It then follows after 
some algebra that Eq. (\ref{equate}) is equivalent to 
\begin{equation}
\label{full}
\left( \frac{2+3f_{NL}}{1+3f_{NL}} \right) (1-n_s) = -2n_t+
\frac{2}{1+3f_{NL}} \frac{\alpha_t}{n_t}  .
\end{equation}

On the other hand, we may also define a  
`spectral index' for the non-linearity parameter, 
$n_{NL} \equiv d \ln f_{NL}/d\ln k$ \cite{chenrunning}. 
In this case, it follows that 
\begin{equation}
\label{indexNG}
s= - \frac{3}{2} \left( \frac{f_{NL}}{1+3f_{NL}} \right) n_{NL}
\end{equation}
and substituting Eqs. (\ref{defindices}), (\ref{fnlcs}) and (\ref{indexNG})
into Eq. (\ref{equate}) then implies that 
\begin{equation}
\label{indexconsistent}
1-n_s= -2n_t +\frac{n_{NL}}{1+3f_{NL}}  .
\end{equation}

Although Eqs. (\ref{full}) and (\ref{indexconsistent}) directly
relate observable parameters, it is not clear whether future observations 
will be able to measure $\alpha_t$ or $n_{NL}$ to sufficient 
accuracy for either of these expressions to be tested.  
Nonetheless, the form of these constraints 
is such that the terms involving $\alpha_t$ and $n_{NL}$ rapidly become 
negligible when $f_{NL} \gg 1$. Indeed, for a detectable non-Gaussianity of 
$f_{NL} >5$, a good approximation to either Eq. (\ref{full}) 
or Eq. (\ref{indexconsistent}) is 
\begin{equation}
\label{indicesconsistent}
1-n_s \simeq -2n_t  .
\end{equation}
Moreover, the consistency equation (\ref{departure}) may be employed to 
express this constraint in terms of the observable parameters 
$\{n_s, r,f_{NL} \}$: 
\begin{equation}
\label{mainresult}
1-n_s \simeq 0.4r \sqrt{f_{NL}}  .
\end{equation}
Eq. (\ref{mainresult}) may 
be interpreted as a new consistency equation 
for brane inflation which is valid when the level of non-Gaussianity 
is sufficiently large and when the warp factor of the throat is 
approximately constant. 

A number of results may be deduced from Eq. (\ref{mainresult}). 
Firstly, it predicts a red density perturbation spectrum $(n_s <1)$ for an 
arbitrary inflaton potential. Such a spectrum 
is favoured by the WMAP3 data 
\cite{spergel}. Secondly, since the observed deviations from a scale-invariant 
spectrum are small, it also predicts that a large non-Gaussian contribution 
to the curvature perturbation will be accompanied by a small 
tensor-scalar ratio, and vice-versa. As a result, 
we should not expect to simultaneously observe both a sizeable non-Gaussian 
signal and a gravitational wave background in these models. 
On the other hand, if a departure from pure scale-invariance is confirmed, 
the present-day upper limit on $f_{NL}$ will impose a {\em lower} 
limit on $r$, and vice-versa.   

In view of this, let us consider what can be deduced from present-day 
observational limits. If a power-law spectrum is assumed, the WMAP3 data 
implies that \cite{spergel}
\begin{equation}
\label{observedlimit}
n_s=0.987^{+0.019}_{-0.037} 
\end{equation}
when the tensor-scalar ratio is included as a free parameter. 
Substituting the upper bound $1-n_s < 0.05$ into Eq. (\ref{mainresult}) 
then yields the limit    
\begin{equation}
\label{boundone}
r < \frac{1}{8\sqrt{f_{NL}}}  .
\end{equation}
Alternatively, imposing the WMAP3 upper bound
$f_{NL} \simeq 300$ implies that
\begin{equation}
\label{boundtwo}
r> \frac{1-n_s}{7}   .
\end{equation}

\begin{figure}[!t]

\includegraphics[width=9cm]{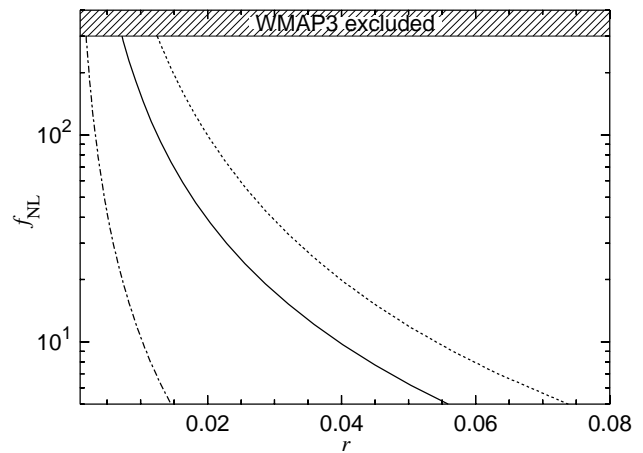}

\caption[] {\label{fig3brane}
Illustrating the constraint equation (\ref{mainresult})
for the WMAP3 best-fit value $n_s=0.987$ (dot-dashed line)
and lower limit $n_s=0.95$ (solid line). The region of parameter 
space consistent with WMAP3 lies below the solid line.
The dashed line represents 
the curve from Fig. \ref{fig2brane}, above which the general 
consistency condition (\ref{departure}) will be detectable. 
}
\end{figure}

Thus, for given values of  $\{ n_s , f_{NL} \}$, the tensor-scalar 
ratio is bounded from above and below. This has a number of important 
implications. In Fig. \ref{fig3brane}, the solid and dot-dashed lines 
represent Eq. (\ref{mainresult}) when the spectral index takes the WMAP3 
lower limit $(n_s =0.95)$ and central value $(n_s=0.987)$, respectively. 
For a given measurement of $f_{NL}$, the allowed values of $r$ 
fall below the solid line. We conclude, therefore, that if Planck  
detects a non-Gaussian signal ($f_{NL} \gsim 5$), 
the tensor-scalar ratio is predicted to lie  
below $r \lsim 0.05$, which is close to the optimal sensitivity 
of Planck. This suggests that a simultaneous detection of both  
non-Gaussianity and gravitational waves by Planck 
will not be possible for these brane inflation scenarios. 
Consequently, a detection of both effects would rule out such models. 

On the other hand, Eq. (\ref{mainresult}) is such 
that the departure away from scale-invariance 
is determined by the combination $r \sqrt{f_{NL}}$. 
This implies that for fixed values of the non-linearity 
parameter in the range $5< f_{NL} < 300$,
the allowed range of $r$ becomes progressively 
narrower as $n_s \rightarrow 1$. For example, 
the relative steepness of the 
dot-dashed line in Fig. \ref{fig3brane} implies that if (as expected) 
the spectral index is ultimately measured to a very high accuracy, a
confirmed detection of the non-linearity parameter (regardless
of the error in the measurement) will strongly 
constrain the tensor-scalar ratio \footnote{We are assuming implicitly here 
that the primordial origin of the three-point correlator of the 
curvature perturbation has been confirmed.}. Indeed,   
for the WMAP3 central value $n_s = 0.987$, 
the tensor-scalar ratio is bounded by $0.002 <  r < 0.015$. 
The existence of such a lower limit on $r$  
is of particular interest given that future 
space-based CMB polarization experiments 
with current detector technology and full sky coverage should be 
able to detect $r \approx  10^{-3}$ at the $3$-sigma level
\cite{SK,vpj}. 

Finally, we have also superimposed Fig. \ref{fig1brane} 
as the dashed line in Fig. 2. 
The general consistency equation (\ref{departure}) will 
only be observable with future all-sky CMB experiments in the region 
above this line and, since it lies in the region of
parameter space that is inconsistent with the 
WMAP3 upper limit (\ref{boundone}), a detection of primordial 
non-Gaussianity by Planck will immediately imply that 
the tensor-scalar ratio will be too small for 
the consistency equation (\ref{departure})
to be detectable. Consequently, an observed violation of the 
standard consistency equation would rule out this class of models.  

\section{Discussion}

There exist many compactification solutions in string theory 
and it is therefore important to identify observational constraints 
and tests of string-theoretic inflation that are insensitive 
to our universe's location in the string landscape. In this paper, 
we have considered this question within the context 
of possible non-Gaussian and gravitational wave signatures   
from brane inflationary scenarios, both for an arbitrary warp factor 
and for the case where the warp factor is approximately constant. 
Furthermore, the form of the inflaton potential has remained unspecified 
throughout. The constraints we have discussed are also independent 
of whether the brane is moving into or out of the throat and therefore apply
to both the UV and IR versions of DBI inflation 
\cite{silverstein,chensolo,shandtye,tipinflation}.  
However, we have only considered radial motion of the brane in the throat, 
and have ignored fluctuations of the brane in its internal 
angular directions. As was emphasized recently \cite{lythriotto}, 
such fluctuations may generate a significant 
contribution to the curvature perturbation. 

In general, the standard inflationary consistency equation 
is violated in brane inflation. The magnitude of the correction 
depends on the non-linearity parameter, $f_{NL}$. 
Consequently, the level of non-Gaussianity  
determines whether or not such a violation will be detectable in
future CMB polarization experiments. We find that 
{\em if} Planck detects a gravitational wave background and 
measures $f_{NL} > 10$, a future all-sky CMB polarization 
satellite should be sufficiently accurate to 
detect (or rule out) the consistency condition (\ref{departure}). 
Similarly, this equation should be testable if Planck  
detects a non-Gaussian signal ($f_{NL} > 5$) and Clover  
measures the tensor-scalar ratio $(r>0.01)$. On the other hand, no 
departure from the standard consistency 
equation will be detectable if Clover fails to detect 
gravitational waves ($r < 0.01$). 

We have identified a new consistency condition, Eq. 
(\ref{mainresult}), which involves the set of observables 
$\{ n_s, r, f_{NL} \}$. This equation applies in the 
limit where the warp factor of the throat is approximately constant, which 
is the case for the KS background in the region near to the tip of the 
throat. This is a potentially strong constraint, 
since it allows general conclusions to be drawn for this 
class of models. In particular, it predicts 
a red scalar perturbation spectrum, $n_s<1$. 

Eq. (\ref{mainresult}) also provides insight into why it has proved difficult 
to construct brane inflation models with both a high 
tensor-scalar ratio and a large non-Gaussian contribution 
to the curvature perturbation \cite{shandtye,tipinflation}. Indeed, the bound 
(\ref{boundtwo}) implies that any model which predicts a very low 
gravitational wave amplitude ($r < 10^{-3}$)
will generate a non-Gaussianity in excess of the 
WMAP3 limit unless the spectral index is extremely close to unity. 
Consequently, the constraint (\ref{boundtwo}) may be employed to 
rule out specific models without needing to explicitly calculate  
the level of non-Gaussianity. 

As an example, let us consider the scenario discussed recently by Panda 
{\em et al.} \cite{panda}, where inflation is generated by the radial 
motion of a probe BPS ${\rm D}3$-brane in the presence of 
a stack of coincident and static BPS ${\rm D5}$-branes. 
The effective DBI action for the 
${\rm D3}$-brane is given by Eq. (\ref{defP}) with 
$f_2 (\phi ) = 2$ and $f_3 =0$ \cite{kutasov}. 
In this model, the warp factor $f_1 (\phi) \rightarrow 1$ 
as $\phi \rightarrow \infty$ and it can be shown that in 
this limit the spectral index and 
tensor-scalar ratio are given by  
\begin{equation}
\label{pandalimit}
1-n_s =\frac{3}{2N_e} , \qquad r= \frac{\sqrt{\sigma}}{N^{3/2}_e}  ,
\end{equation}
respectively, where $N_e$ is the number of e-folds before the 
end of inflation and $\sigma \ll 1$ is a parameter that  
depends on the number of ${\rm D5}$-branes that are present \cite{panda}. 
The constraint (\ref{boundtwo}) then implies that 
the WMAP3 non-Gaussianity limit is only satisfied 
if
$N_e < 22 \sigma$,
but even for $\sigma = 1$ this would require  
$(1-n_s) >0.07$, which is
marginally inconsistent 
with observations. Hence, inflation in this region of parameter space 
is probably ruled out. 

Finally, Eq. (\ref{mainresult}) implies a {\em lower} 
limit on the tensor-scalar ratio for a given value of the spectral index.  
In general, we expect $(1-n_s)$ to be of the order a 
few percent since typically $1-n_s \approx 1/N_e$, where $30< N_e <60$. 
In this case, a detection of non-Gaussianity by Planck would  
imply that $r >10^{-3}$, which will be accessible to the next 
generation of CMB polarization experiments \cite{SK,vpj}. This 
is interesting since the potential benefits of building 
such an experiment have been questioned recently given the dearth 
of well-motivated models which predict $10^{-3} < r < 10^{-2}$ 
\cite{desert}. Our results therefore provide strong motivation 
for developing specific brane inflation models
based on more general throat geometries and compactification schemes. 

\section*{Acknowledgments}
We thank Y-S. Song and L. Knox \cite{SK} 
for making the numerical results of their paper 
available to us. We also thank D. Lyth for organising the 
workshop on non-Gaussian perturbations where 
this work was initiated.

\end{document}